\documentclass[aps,prl,preprint,tightenlines,superscriptaddress,showpacs,byrevtex]{revtex4}
\usepackage{longtable}
\usepackage{amsmath}
\usepackage{graphicx}
\usepackage{dcolumn}

\newcommand{\mm}{M_{\rm miss}}
\newcommand{\de}{\Delta E}
\newcommand{\up}{\Upsilon}
\newcommand{\upns}{\up(nS)}
\newcommand{\upf}{\up(5S)}
\newcommand{\pipi}{\pi^+\pi^-}
\newcommand{\pnpn}{\pi^0\pi^0}

\newcommand{\br}{{\cal B}}

\newcommand{\ee}{e^+e^-}
\newcommand{\mumu}{\mu^+\mu^-}
\newcommand{\lplm}{l^+l^-}
\newcommand{\mev}{{\rm MeV}}
\newcommand{\mevc}{{\rm MeV/c}}
\newcommand{\mevcc}{{\rm MeV/c^2}}
\newcommand{\gevcc}{{\rm GeV/c^2}}
\newcommand{\zb}{Z_b}
\newcommand{\zbn}{Z_b^0}
\newcommand{\zbnf}{\zbn(10610)}
\newcommand{\fix}{(fixed)}
\newcommand{\tl}{2\log {\mathcal L}}
\newcommand{\dtl}{\Delta(-\tl)}
\newcommand{\anr}{A_{\rm nr}}
\newcommand{\phinr}{\phi_{\rm nr}}
\newcommand{\fsig}{f_{\rm sig}}

\begin{document}

\preprint{
\vbox{ 
\hbox{   }
                 \hbox{BELLE-CONF-1271}
}
}

\title{\boldmath Evidence for a $\zbnf$ in Dalitz analysis of $\upf\to\upns\pnpn$}


\begin{abstract}
  We report the first observation of $\upf\to\up(1,2S)\pnpn$ decays.
  Evidence for the $\zbnf$ with $4.9\sigma$ significance is 
  found in a Dalitz plot analysis of $\upf\to\up(2S)\pnpn$ decays.
  The results are obtained with a $121.4\,{\rm fb}^{-1}$ data sample collected
  with the Belle detector at the $\upf$ resonance at the KEKB
  asymmetric-energy $\ee$ collider.

\end{abstract}


\affiliation{University of Bonn, Bonn}
\affiliation{Budker Institute of Nuclear Physics SB RAS and Novosibirsk State University, Novosibirsk 630090}
\affiliation{Faculty of Mathematics and Physics, Charles University, Prague}
\affiliation{Chiba University, Chiba}
\affiliation{University of Cincinnati, Cincinnati, Ohio 45221}
\affiliation{Department of Physics, Fu Jen Catholic University, Taipei}
\affiliation{Justus-Liebig-Universit\"at Gie\ss{}en, Gie\ss{}en}
\affiliation{Gifu University, Gifu}
\affiliation{The Graduate University for Advanced Studies, Hayama}
\affiliation{Gyeongsang National University, Chinju}
\affiliation{Hanyang University, Seoul}
\affiliation{University of Hawaii, Honolulu, Hawaii 96822}
\affiliation{High Energy Accelerator Research Organization (KEK), Tsukuba}
\affiliation{Hiroshima Institute of Technology, Hiroshima}
\affiliation{University of Illinois at Urbana-Champaign, Urbana, Illinois 61801}
\affiliation{Indian Institute of Technology Guwahati, Guwahati}
\affiliation{Indian Institute of Technology Madras, Madras}
\affiliation{Indiana University, Bloomington, Indiana 47408}
\affiliation{Institute of High Energy Physics, Chinese Academy of Sciences, Beijing}
\affiliation{Institute of High Energy Physics, Vienna}
\affiliation{Institute of High Energy Physics, Protvino}
\affiliation{Institute of Mathematical Sciences, Chennai}
\affiliation{INFN - Sezione di Torino, Torino}
\affiliation{Institute for Theoretical and Experimental Physics, Moscow}
\affiliation{J. Stefan Institute, Ljubljana}
\affiliation{Kanagawa University, Yokohama}
\affiliation{Institut f\"ur Experimentelle Kernphysik, Karlsruher Institut f\"ur Technologie, Karlsruhe}
\affiliation{Korea Institute of Science and Technology Information, Daejeon}
\affiliation{Korea University, Seoul}
\affiliation{Kyoto University, Kyoto}
\affiliation{Kyungpook National University, Taegu}
\affiliation{\'Ecole Polytechnique F\'ed\'erale de Lausanne (EPFL), Lausanne}
\affiliation{Faculty of Mathematics and Physics, University of Ljubljana, Ljubljana}
\affiliation{Luther College, Decorah, Iowa 52101}
\affiliation{University of Maribor, Maribor}
\affiliation{Max-Planck-Institut f\"ur Physik, M\"unchen}
\affiliation{University of Melbourne, School of Physics, Victoria 3010}
\affiliation{Graduate School of Science, Nagoya University, Nagoya}
\affiliation{Kobayashi-Maskawa Institute, Nagoya University, Nagoya}
\affiliation{Nara University of Education, Nara}
\affiliation{Nara Women's University, Nara}
\affiliation{National Central University, Chung-li}
\affiliation{National United University, Miao Li}
\affiliation{Department of Physics, National Taiwan University, Taipei}
\affiliation{H. Niewodniczanski Institute of Nuclear Physics, Krakow}
\affiliation{Nippon Dental University, Niigata}
\affiliation{Niigata University, Niigata}
\affiliation{University of Nova Gorica, Nova Gorica}
\affiliation{Osaka City University, Osaka}
\affiliation{Osaka University, Osaka}
\affiliation{Pacific Northwest National Laboratory, Richland, Washington 99352}
\affiliation{Panjab University, Chandigarh}
\affiliation{Peking University, Beijing}
\affiliation{Princeton University, Princeton, New Jersey 08544}
\affiliation{Research Center for Electron Photon Science, Tohoku University, Sendai}
\affiliation{Research Center for Nuclear Physics, Osaka University, Osaka}
\affiliation{RIKEN BNL Research Center, Upton, New York 11973}
\affiliation{Saga University, Saga}
\affiliation{University of Science and Technology of China, Hefei}
\affiliation{Seoul National University, Seoul}
\affiliation{Shinshu University, Nagano}
\affiliation{Sungkyunkwan University, Suwon}
\affiliation{School of Physics, University of Sydney, NSW 2006}
\affiliation{Tata Institute of Fundamental Research, Mumbai}
\affiliation{Excellence Cluster Universe, Technische Universit\"at M\"unchen, Garching}
\affiliation{Toho University, Funabashi}
\affiliation{Tohoku Gakuin University, Tagajo}
\affiliation{Tohoku University, Sendai}
\affiliation{Department of Physics, University of Tokyo, Tokyo}
\affiliation{Tokyo Institute of Technology, Tokyo}
\affiliation{Tokyo Metropolitan University, Tokyo}
\affiliation{Tokyo University of Agriculture and Technology, Tokyo}
\affiliation{Toyama National College of Maritime Technology, Toyama}
\affiliation{CNP, Virginia Polytechnic Institute and State University, Blacksburg, Virginia 24061}
\affiliation{Wayne State University, Detroit, Michigan 48202}
\affiliation{Yamagata University, Yamagata}
\affiliation{Yonsei University, Seoul}
  \author{I.~Adachi}\affiliation{High Energy Accelerator Research Organization (KEK), Tsukuba} 
  \author{K.~Adamczyk}\affiliation{H. Niewodniczanski Institute of Nuclear Physics, Krakow} 
  \author{H.~Aihara}\affiliation{Department of Physics, University of Tokyo, Tokyo} 
  \author{K.~Arinstein}\affiliation{Budker Institute of Nuclear Physics SB RAS and Novosibirsk State University, Novosibirsk 630090} 
  \author{Y.~Arita}\affiliation{Graduate School of Science, Nagoya University, Nagoya} 
  \author{D.~M.~Asner}\affiliation{Pacific Northwest National Laboratory, Richland, Washington 99352} 
  \author{T.~Aso}\affiliation{Toyama National College of Maritime Technology, Toyama} 
  \author{V.~Aulchenko}\affiliation{Budker Institute of Nuclear Physics SB RAS and Novosibirsk State University, Novosibirsk 630090} 
  \author{T.~Aushev}\affiliation{Institute for Theoretical and Experimental Physics, Moscow} 
  \author{T.~Aziz}\affiliation{Tata Institute of Fundamental Research, Mumbai} 
% ...
\collaboration{The Belle Collaboration}


\maketitle

\section{Introduction}
Recently the Belle Collaboration reported the observation of two 
narrow structures in $\pi^\pm\upns$ invariant mass in the
$\upf\to\upns\pipi$ decays ($n=1,2,3$)~\cite{zb_paper}.
The measured masses and widths of the two structures are
$M_1=10607.2\pm 2.0\ \mevcc$, $\Gamma_1=18.4\pm 2.4\ \mev$ and
$M_2=10652.2\pm 1.5\ \mevcc$, $\Gamma_2=11.5\pm 2.2\ \mev$, respectively. 
Angular analysis suggests that these states have 
$I^G(J^P)=1^+(1^+)$ quantum numbers~\cite{zb_helicity}.
The measured masses are a few $\mevcc$ above the thresholds for the open
beauty channels $B^*\bar{B}$ $(10604.6\ \mevcc)$  and $B^*\bar{B}$ 
$(10604.6\ \mevcc)$ suggesting a ``molecular'' nature for these states, 
which is consistent with many of their observed properties~\cite{zb_molecular}.
This observation motivates us to search for a neutral partner of 
these states in the resonant substructure of
$\upf\to\upns\pnpn$ decays.

\section{Selection Criteria}
We use a $121.4\pm 1.9\ {\rm fb}^{-1}$ data sample collected on the peak
of the $\upf$ resonance with the Belle detector~\cite{belle} at the KEKB 
asymmetric energy $\ee$ collider~\cite{kekb}.
The Belle detector is a large-solid-angle magnetic spectrometer that
consists of a silicon vertex detector, a central drift chamber, an array
of aerogel threshold Cherenkov counters, a barrel-like arrangement of
time-of-flight scintillation counters, and an electromagnetic calorimeter
comprised of CsI(Tl) crystals located inside a superconducting solenoid
that provides a 1.5~T magnetic field. An iron flux-return located outside
the coil is instrumented to detect $K_L^0$ mesons and to identify muons.
The detector is described in detail elsewhere~\cite{belle}.

$\upf$ candidates are formed from $\upns\pnpn$ ($n=1,2$) 
combination.
We reconstruct $\upns$ candidates from pairs of leptons ($\ee$ and $\mumu$)
with invariant mass in the range from 8 to 11~$\gevcc$.
An additional decay channel is used for the $\up(2S)$: 
$\up(2S)\to\up(1S)[l^+l^-]\pipi$.
Charged tracks are required to have transverse momentum, $p_t>$,
greater than $50\,\mevc$. 
We also impose a requirement on the impact parameters:
$dr<0.3$~cm and $|dz|<2.0$~cm, where $dr$ and $dz$ are the impact 
parameters in the $r$-$\phi$ and longitudinal directions, respectively.
Muon and electron candidates are required to be positively identified.
No requirement on the particle identification is used for the pions.
Candidate $\pi^0$ mesons are selected from pairs of photons with an 
invariant mass within 15~$\mevcc$ of the nominal $\pi^0$ mass.
Energy greater than 50 (75) MeV is required for each photon in the 
barrel (endcap).
We use the quality of the $\pi^0$ mass-constrained fits to suppress the 
background; the sum of $\chi^2(\pi^0_1)+\chi^2(\pi^0_2)$ is required to be less than 
20 (10) for the $\upns\to\mumu$, $\up(1S)\pipi$ ($\upns\to\ee$).

We use the energy difference, $\de= E_{\rm cand}-E_{\rm CM}$, and momentum 
$P$ to suppress background, where $E_{\rm cand}$ and $P$ are the
energy and momentum of the reconstructed $\upf$ candidate in the 
center-of-mass (c.m.) frame, and $E_{\rm CM}$ is the c.m. energy of the 
two beams. 
$\upf$ candidates are required to satisfy the requirements
$-0.2<\de<0.14$~GeV and $P<0.2$~GeV$/$c.
The large potential background from QED processes such as $\ee\to\lplm (n)\gamma$ 
is suppressed using the missing mass associated with the 
$\lplm$ system, $\mm(\lplm)$, 
calculated as $\mm(\lplm)=\sqrt{(E_{CM}-E_{\lplm})^2-P_{\lplm}^2}$,
where $E_{\lplm}$ and $P_{\lplm}$ are the energy and momentum of the 
$\lplm$ system measured in the c.m. frame. We require 
$\mm(\lplm)>0.15\,(0.30)$~$\gevcc$ for the $\upns\to\mumu$ ($\ee$).
We select the candidate with the smallest $\chi^2(\pi^0_1)+\chi^2(\pi^0_2)$ 
in the rare cases (1-2\%) in which there is more than one candidate 
in the event.

\begin{figure*}
  \includegraphics[width=0.32\textwidth] {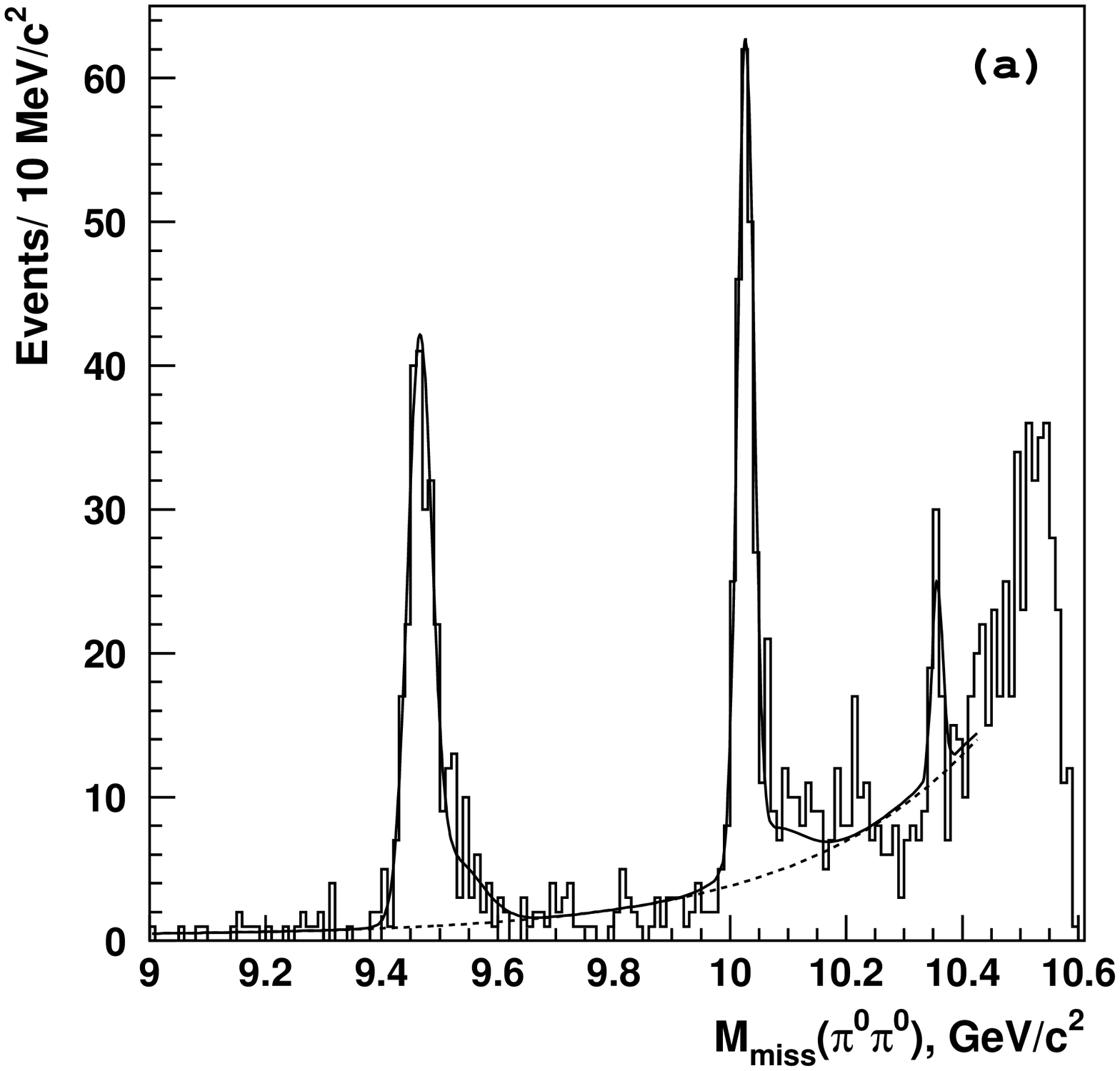}
  \includegraphics[width=0.32\textwidth] {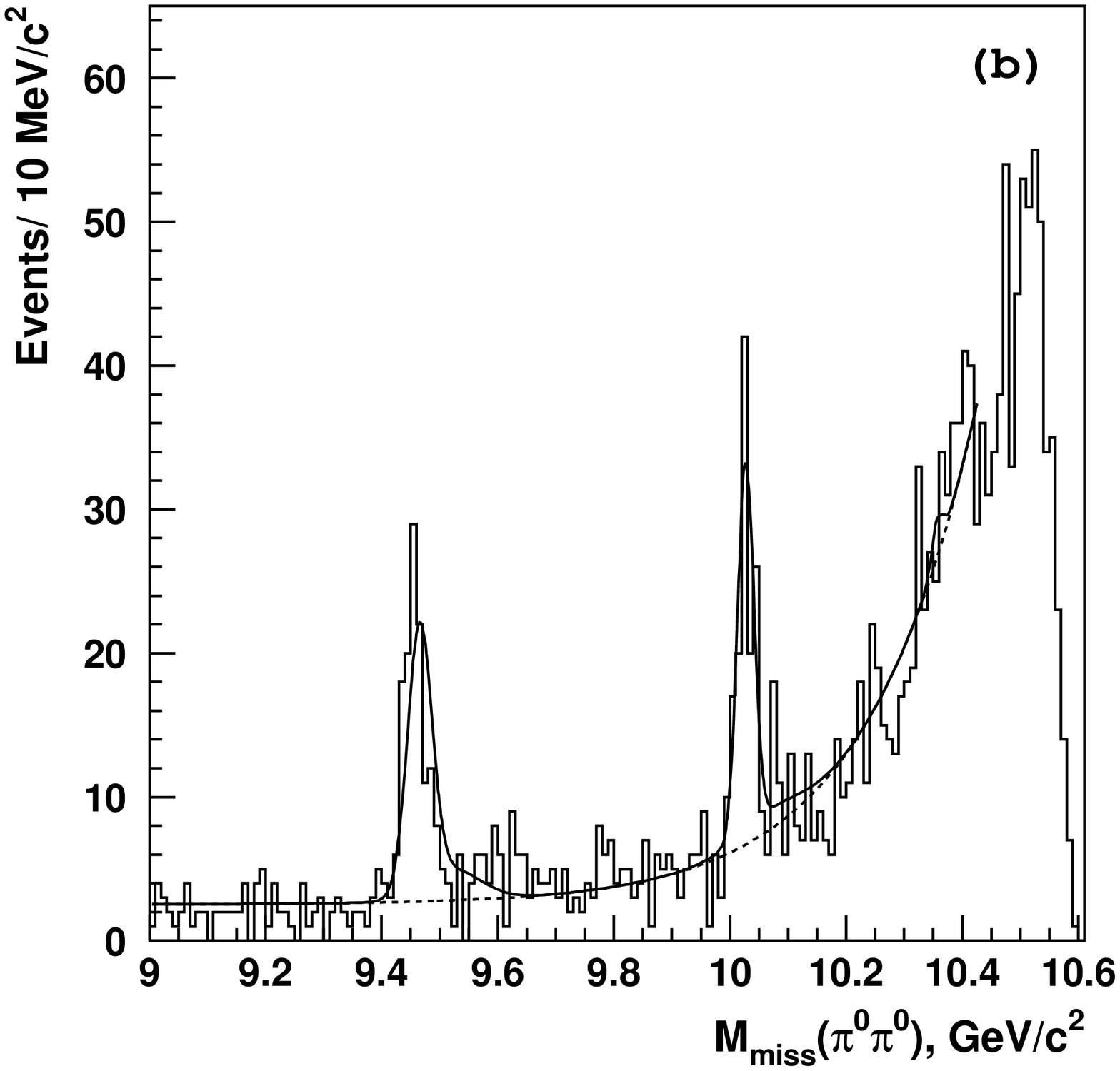}
  \includegraphics[width=0.32\textwidth] {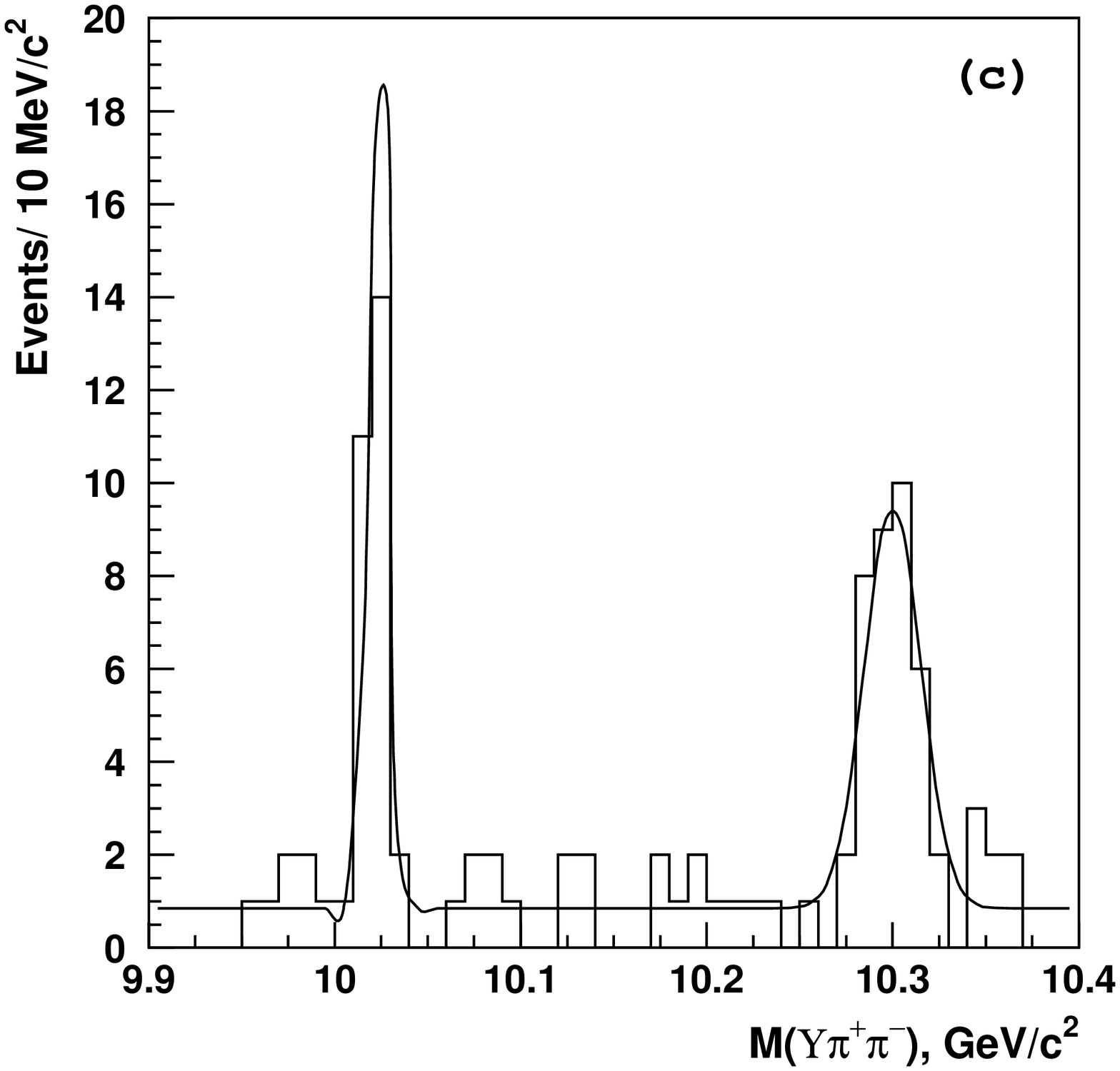}
  \caption{The $\pnpn$ missing mass distribution for
    $\upns\pnpn$, (a) $\upns\to\mumu$ and (b) $\upns\to\ee$ candidates.
    The $M(\up(1S)\pipi)$ distribution for $\up(2S)\to\up(1S)\pipi$ 
    candidates is shown in (c).
    Histograms represent the data. The solid curves show the fit result 
    while the dashed curves correspond to the background contributions.}
  \label{fig:mmpnpn}
\end{figure*}

$\upf\to\upns[\lplm]\pnpn$ candidates are identified via the missing mass 
recoiling against the $\pnpn$ system, $\mm(\pnpn)$.
Figures \ref{fig:mmpnpn}~(a) and \ref{fig:mmpnpn}~(b) show the 
$\mm(\pnpn)$ distributions 
for $\upf\to\upns[\lplm]\pnpn$ candidates.
We fit these distributions to extract the $\upns$ signal yield.
The signal probability density function (PDF) is described by a sum of two 
Gaussians for each $\upns$ resonance with parameters fixed from the signal 
Monte Carlo (MC) sample. The background PDF is parameterized
by the sum of constant and exponential functions.

For $\upf\to\up(2S)[\up(1S)\pipi]\pnpn$ decays, $\up(1S)$ candidates 
are selected from $l^+l^-$ pairs with invariant mass within 
$150\,\mevcc$ of the nominal $\up(1S)$ mass.
A mass-constrained fit is used for $\up(1S)$ candidates to improve 
the momentum resolution.
We apply the same requirements on $\de$ and $P$ described above 
for reconstructed $\upf$ candidate.  We use the invariant mass
of $\up(1S)\pipi$ to select the signal candidates.
Figure~\ref{fig:mmpnpn}~(c) shows the $M(\up(1S)\pipi)$ distribution 
for the $[\up(1S)\pipi]\pnpn$ events.
We fit this distribution to extract the $\up(2S)[\up(1S)\pipi]$ 
signal yield. The signal PDF is described by a Gaussian function with 
parameters fixed from signal MC. The background PDF is described by a 
constant. The cross-feed from the decay $\upf\to\up(2S)[\up(1S)\pnpn]\pipi$ 
contributes as a broad peak around $10.3$~$\gevcc$. Its contribution is 
parameterized by a Gaussian function.

Table~\ref{tab:yield} summarizes the definition of the signal region, 
signal yield, MC efficiency, measured branching fraction (only the statistical 
uncertainty is shown), number of selected events and purity.
The reconstruction efficiency is obtained using MC with with the 
$\upns\pnpn$ system distributed uniformly over three-body phase space. 
The branching fraction is calculated by $\br=\frac{N_{\rm sig}}{\epsilon {\cal L}\sigma(\ee\to\upf)}$,
where $N_{\rm sig}$ is number of signal events, $\epsilon$ is reconstruction efficiency, ${\cal L}$ is 
integrated luminosity. We use the value of $\sigma(\ee\to\upf)=0.340\pm 0.016$~nb obtained with
$121.4\,{\rm fb}^{-1}$ data.

\begin{table*}
\caption{Definition of the signal region, signal yield, MC efficiency, 
measured branching fraction, number of selected events and purity.}

\medskip
\label{tab:yield}

\begin{tabular*}{\textwidth}{@{\extracolsep{\fill}}lcccccc}\hline\hline
Final state & Signal region, $\gevcc$ & Signal yield & $\epsilon$, \% &$\br$, $10^{-3}$ & Events & Purity\\ 
\hline
$\up(1S)\to\mumu$ & $9.41<\mm(\pnpn)<9.53$  & $261\pm 15$ & $11.2$ & $2.28\pm 0.13$ & 247 & $0.95$\\
$\up(1S)\to\ee$ &   $9.41<\mm(\pnpn)<9.53$  & $123\pm 13$ & $5.61$ & $2.15\pm 0.23$ & 140 & $0.78$\\
$\up(2S)\to\mumu$ & $9.99<\mm(\pnpn)<10.07$ & $241\pm 18$ & $8.04$ & $3.77\pm 0.28$ & 253 & $0.87$ \\
$\up(2S)\to\ee$ &   $9.99<\mm(\pnpn)<10.07$ & $108\pm 13$ & $3.58$ & $3.84\pm 0.46$ & 151 & $0.66$\\
$\up(2S)\to\up(1S)\pipi$&$10.00<M(\up\pipi)<10.05$& $24\pm 5$  & $2.27$ & $2.85\pm 0.60$ & 28 & $0.86$\\ 
\hline\hline
\end{tabular*}
\end{table*}

The main sources of systematic uncertainty in the branching fraction 
measurement are: uncertainties in the signal and background PDFs 
used in the fit mainly due to data/MC width differences: $5\%$;
uncertainty in the $\ee\to\upf$ cross section: $5\%$;
$\br(\up(nS)\to l^+l^-)$: $2\%$ and $9\%$ for the $\up(1S)$ and 
$\up(2S)$~\cite{PDG};
luminosity: $1.5\%$;
$\pi^0$ reconstruction: $5\%$;
muon identification: $1\%$; electron identification: $3\%$;
tracking: $0.7\%$.
The total systematic errors are $9\%$ for $\up(1S)\pnpn$ 
and $13\%$ for $\up(2S)\pnpn$.
We calculate the weighted average of $\br(\upf\to\up(nS)\pnpn)$ 
in the various $\upns$ decay channels and obtain
$\br(\upf\to\up(1S)\pnpn)=(2.25\pm 0.11\pm 0.20)\times 10^{-3}$ and
$\br(\upf\to\up(2S)\pnpn)=(3.66\pm 0.22\pm 0.48)\times 10^{-3}$.
These are approximately one half of the corresponding values of
$\br(\upf\to\upns\pipi)$~\cite{belle_ypipi},
consistent with expectations from isospin.

\section{Dalitz Analysis}
\begin{figure*}
  \includegraphics[width=0.24\textwidth] {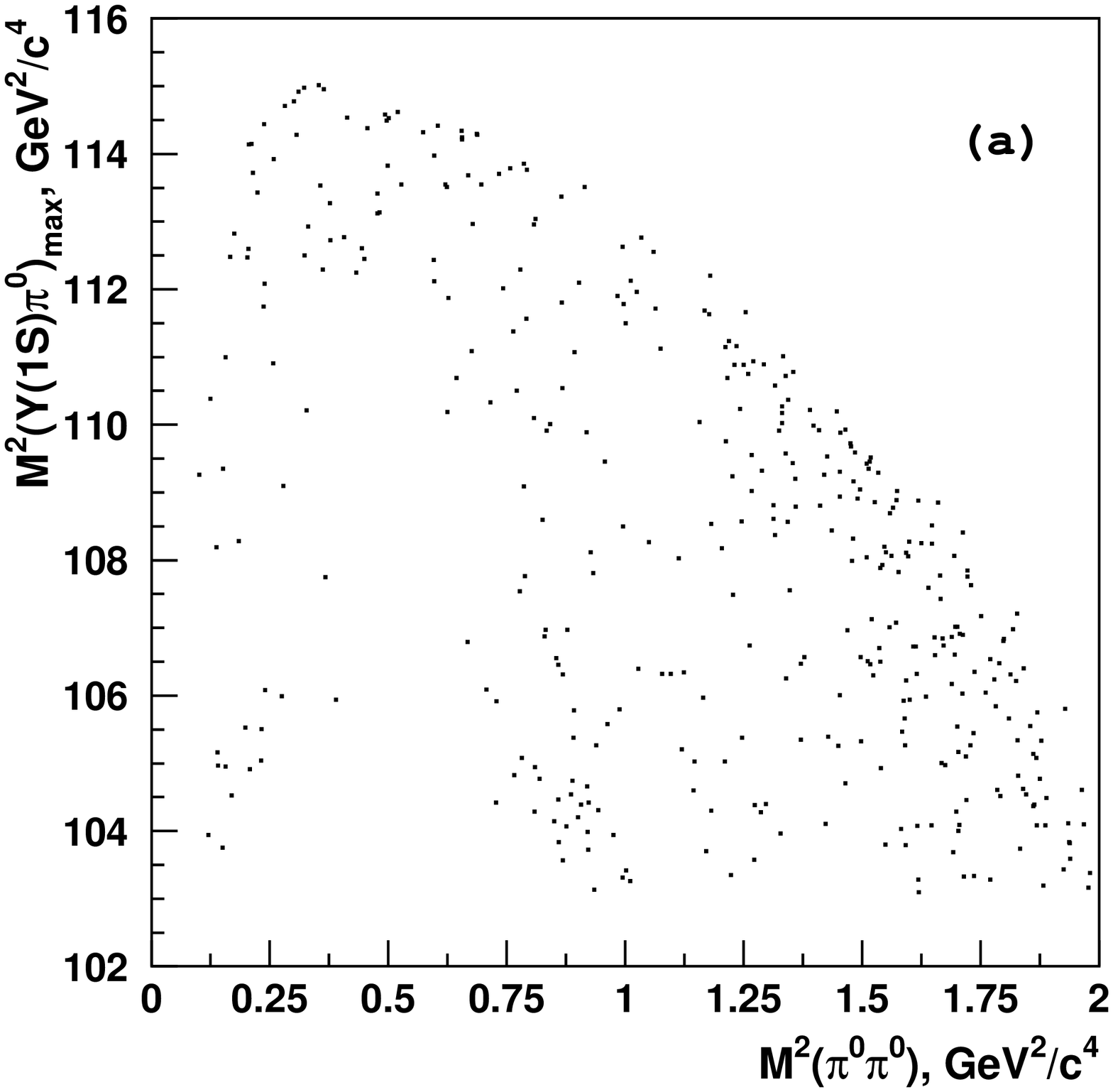}
  \includegraphics[width=0.24\textwidth] {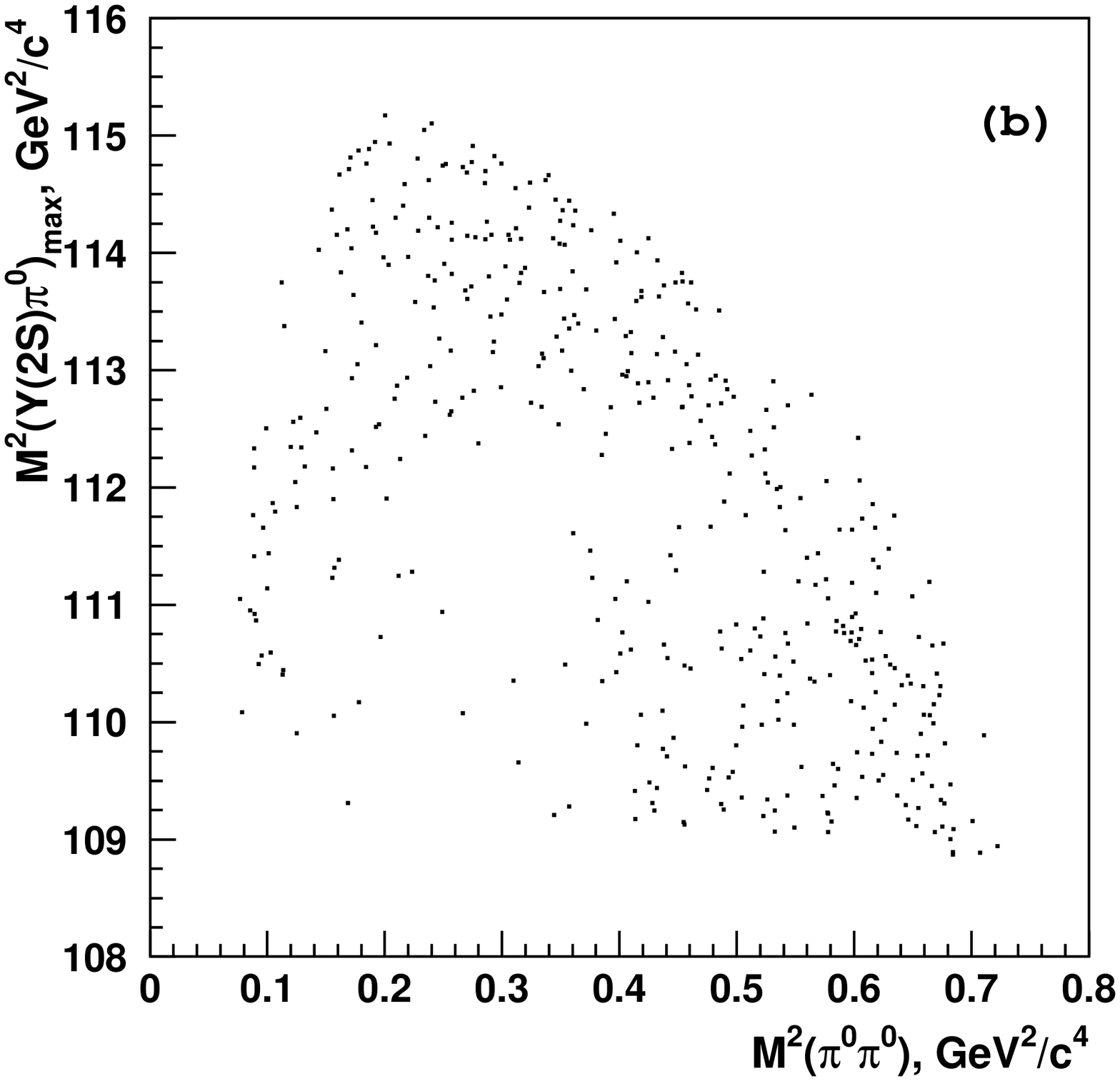}
  \includegraphics[width=0.24\textwidth] {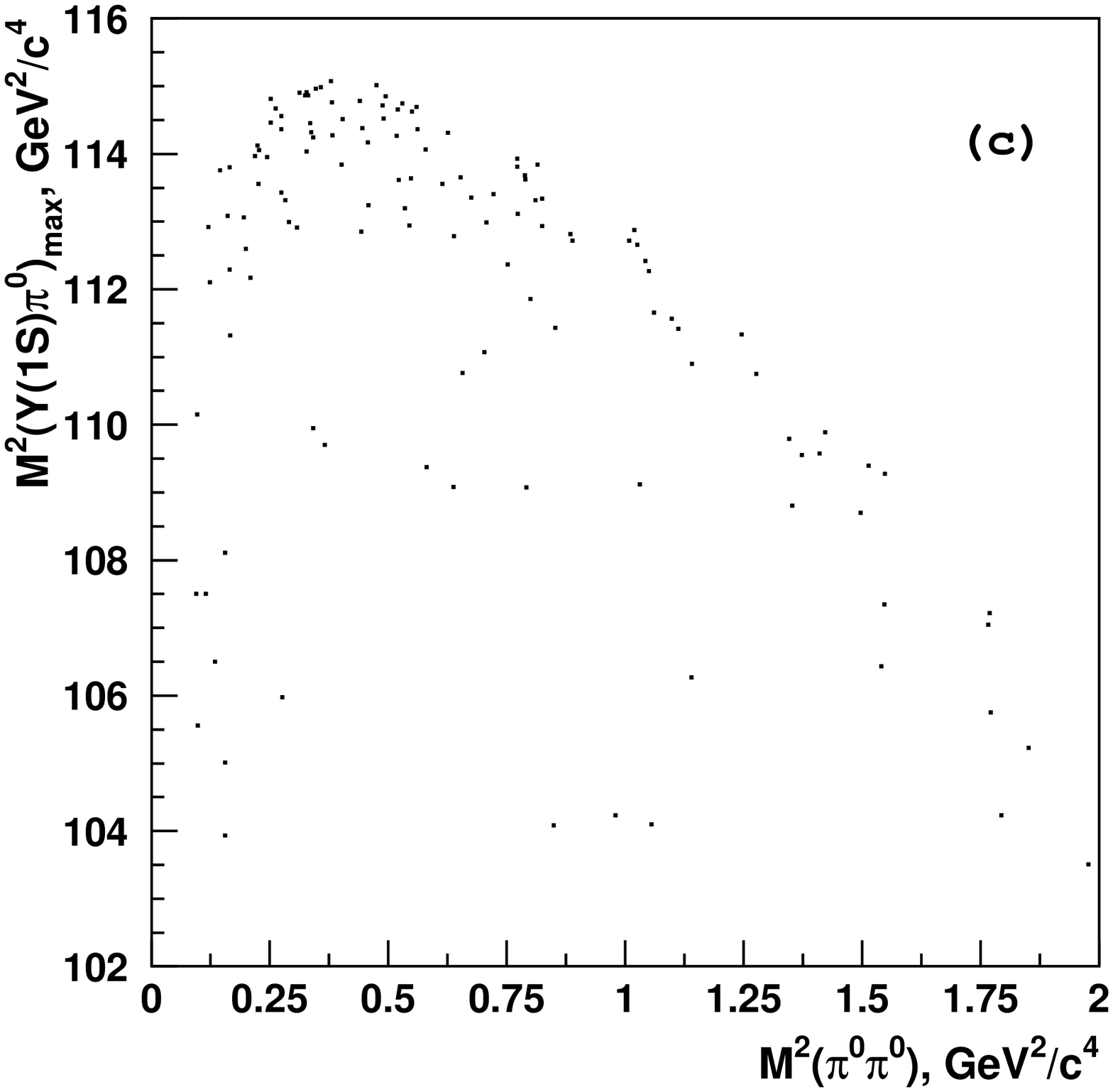}
  \includegraphics[width=0.24\textwidth] {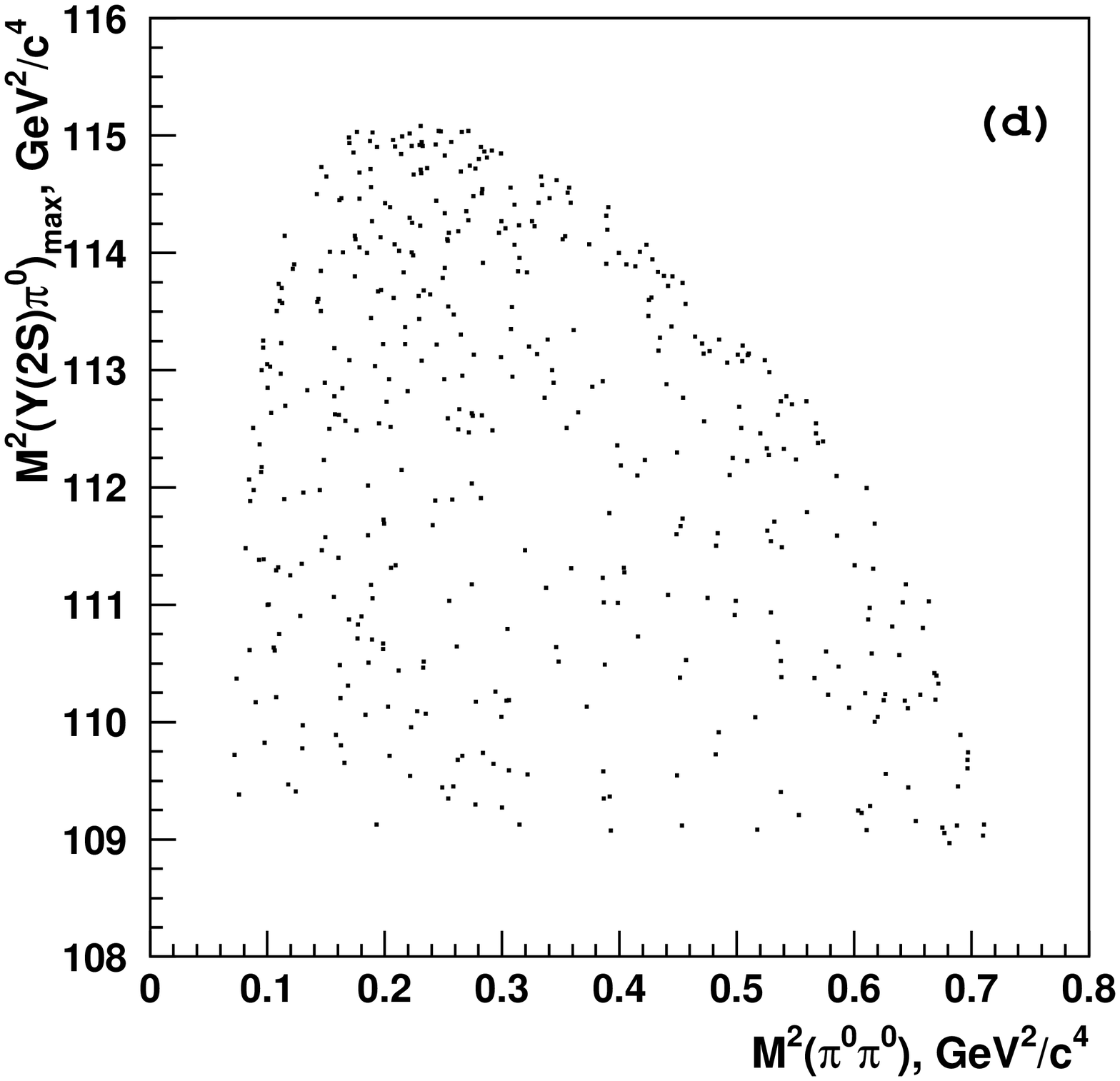}
  \caption{Dalitz plots for selected (a) $\up(1S)\pnpn$, (b) $\up(2S)\pnpn$ candidates.
    Dalitz plots for events in the 
    (c) $\up(1S)\pnpn$, (d) $\up(2S)\pnpn$ sidebands.}
  \label{fig:dalitz}
\end{figure*}

We define the following sideband regions for the study of background:
for the $\up(1S)\pnpn$ final state : $9.20\,\gevcc<\mm(\pnpn)<9.35\,\gevcc$ and 
$9.60\,\gevcc<\mm(\pnpn)<9.75\,\gevcc$;
and for the $\up(2S)\pnpn$ final state: $9.80\,\gevcc<\mm(\pnpn)<9.95\,\gevcc$ and 
$10.15\,\gevcc<\mm(\pnpn)<10.30\,\gevcc$.
Figure~\ref{fig:dalitz} shows the Dalitz plot distributions for the selected
$\upf\to\upns\pnpn$ candidates from the signal region and sidebands. 
Before analyzing Dalitz distributions for events in the signal region, 
we determine the PDF for background.
Samples of background events are selected in $\upns$ mass sidebands and 
then refitted to the nominal mass of the corresponding $\upns$ 
state to match the phase space boundaries. 
We parameterize the background PDF by the following function:
\begin{equation}
1+p_1\exp(-q_1 s_3 + p_2\exp(-q_2 (s_{\rm min}-a_2)),
\end{equation}
where $p_1$, $p_2$, $q_1$ and $q_2$ are fit parameters.
Here $s_3=M^2(\pnpn)$ and 
$s_{\rm min}={\rm Min}(s_1, s_2)$, $s_{1,2}=M^2(\upns\pi^0_{1,2})$.
The kinematical limit $a_2$ is $92$ and $103$~GeV$^2/c^4$ for the $\up(1S)$ and $\up(2S)$, respectively.
Variation of the reconstruction efficiency over the Dalitz plot is 
determined using a large sample of MC with a uniform phase space 
distribution.
We use the following function to parameterize efficiency variations:
\begin{equation}
\epsilon=1+c\{1-e^{-(s_3-a_0)/b_0}\}\{1-e^{-(a_1-s_{\rm max})/b_1}\},
\end{equation}
where $c$, $b_0$ and $b_1$ are fit parameters. 
Here $s_{\rm max}={\rm Max}(s_1, s_2)$.
The parameters $a_0$ and $a_1$ are defined as 
$a_0=4m^2_{\pi^0}$, $a_1 = (m_{\upf}-m_{\pi^0})^2$.

The amplitude analysis of three-body $\upf\to\upns\pnpn$ decays 
uses an unbinned maximum likelihood fit.
We describe the three-body signal amplitude as a sum of quasi-two-body
amplitudes:
\begin{equation}
M(s_1,s_2)=A_{Z1}+A_{Z2}+A_{f_0}+A_{f_2}+\anr\ ,
\end{equation}
where $A_{Z1}$ and $A_{Z2}$ are amplitudes for contributions from the 
$\zbnf$ and $\zbn(10650)$, respectively. 
The amplitudes $A_{f_0}$, $A_{f_2}$ and $\anr$ are the contributions from the 
$\pnpn$ system in an $f_0(980$, $f_2(1275)$ and a non-resonant state, respectively.
Here we assume that the dominant 
contributions are from amplitudes that preserve the orientation of the spin of 
the heavy quarkonium state and thus, both pions in the cascade decay 
$\upf\to\zbn\pi^0\to\upns\pnpn$ are emitted in an $S$-wave with respect
to the heavy quarkonium system. As demonstrated in Ref.~\cite{zb_helicity},
angular analysis supports this assumption. Consequently, we parameterize both
amplitudes with an $S$-wave Breit-Wigner function
\begin{equation}
{\rm BW}(s,M,\Gamma)=\frac{\sqrt{M\Gamma}}{M^2-s-iM\Gamma}\,,
\end{equation}
where we neglect the possible $s$ dependence of the resonance width.
Both amplitudes are symmetrized with respect to $\pi^0$ interchange.
The masses and widths are fixed to the 
values obtained in the $\upns\pipi$ analysis: $M(Z_1)=10607.2\,\mevcc$, 
$\Gamma(Z_1)=18.4\,\mevc$, $M(Z_2)=10652.2\,\mevcc$,
$\Gamma(Z_2)=11.5\,\mevc$~\cite{zb_paper}.
Contributions from the $f_0(980)$ and $f_2(1275)$ are also included in the fit.
We use a Flatt\'{e} function for the $f_0(980)$ and a Breit-Wigner function 
for the $f_2(1275)$.
Coupling constants of the $f_0(980)$ were fixed at values from 
the $B^+\to K^+\pipi$ analysis: $M=950\,\mevcc$, $g_{\pi\pi}=0.23$, 
$g_{KK}=0.73$~\cite{kpipi}. 
The mass and width of the $f_2(1275)$ resonance are fixed to the world average 
values~\cite{PDG}.
Following suggestions in Ref.~\cite{voloshin}, the non-resonant
amplitude $\anr$ is parameterized as 
\begin{equation}
\anr=\anr^1 e^{i\phinr^1} + \anr^2 e^{i\phinr^2} s_3\ ,
\end{equation}
where $\anr^1$, $\anr^2$, $\phinr^1$ and $\phinr^2$ 
are free parameters in the fit.
As there is only sensitivity to the relative amplitudes and
phases between decay modes, we fix $\anr^1=10.0$ and $\phinr^1=0.0$.

The logarithmic likelihood function is defined as
\begin{equation}
{\mathcal L}=-2\sum\log\{\epsilon(s_1,s_2) (\fsig S(s_1,s_2)+(1-\fsig) B(s_1,s_2))\}\ ,
\end{equation}
where $S(s_1,s_2)$ is $|M(s_1,s_2)|^2$ convoluted with the detector resolution
($6.0\,\mevcc$ for $\upns\pi^0$ combinations), $\epsilon(s_1,s_2)$ describes 
the variation of the reconstruction efficiency over the Dalitz plot and 
$\fsig$ is the fraction of signal events in the data sample. The fraction 
$\fsig$ is determined separately for each $\upns$ decay mode 
(see Table~\ref{tab:yield}). The function $B(s_1,s_2)$ describes the 
distribution of background events over the 
phase space. Both products $\epsilon(s_1,s_2)\cdot S(s_1,s_2)$ and 
$\epsilon(s_1,s_2)\cdot B(s_1,s_2)$ are normalized to unity.

\begin{figure*}
  \includegraphics[width=0.32\textwidth] {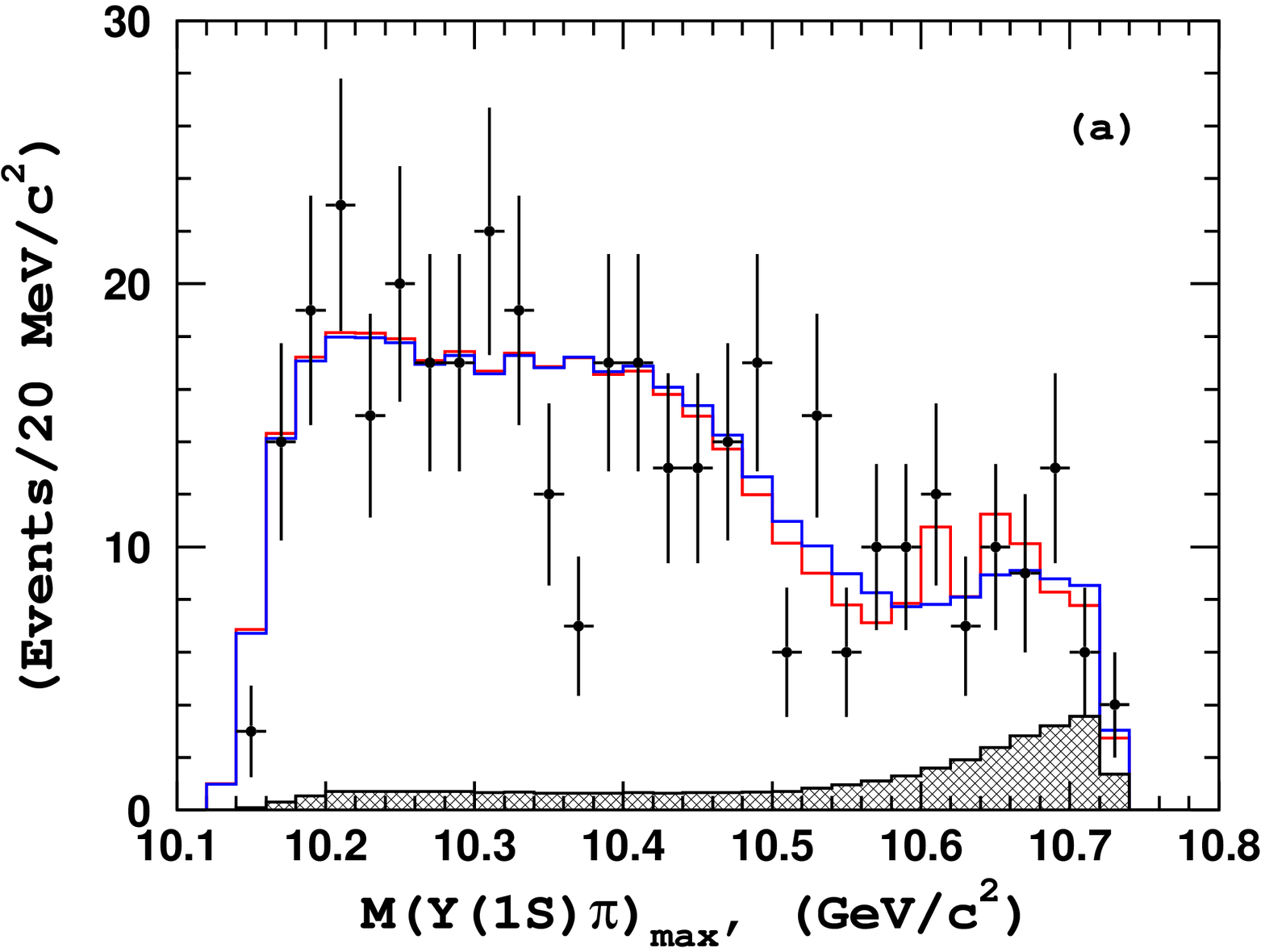}
  \includegraphics[width=0.32\textwidth] {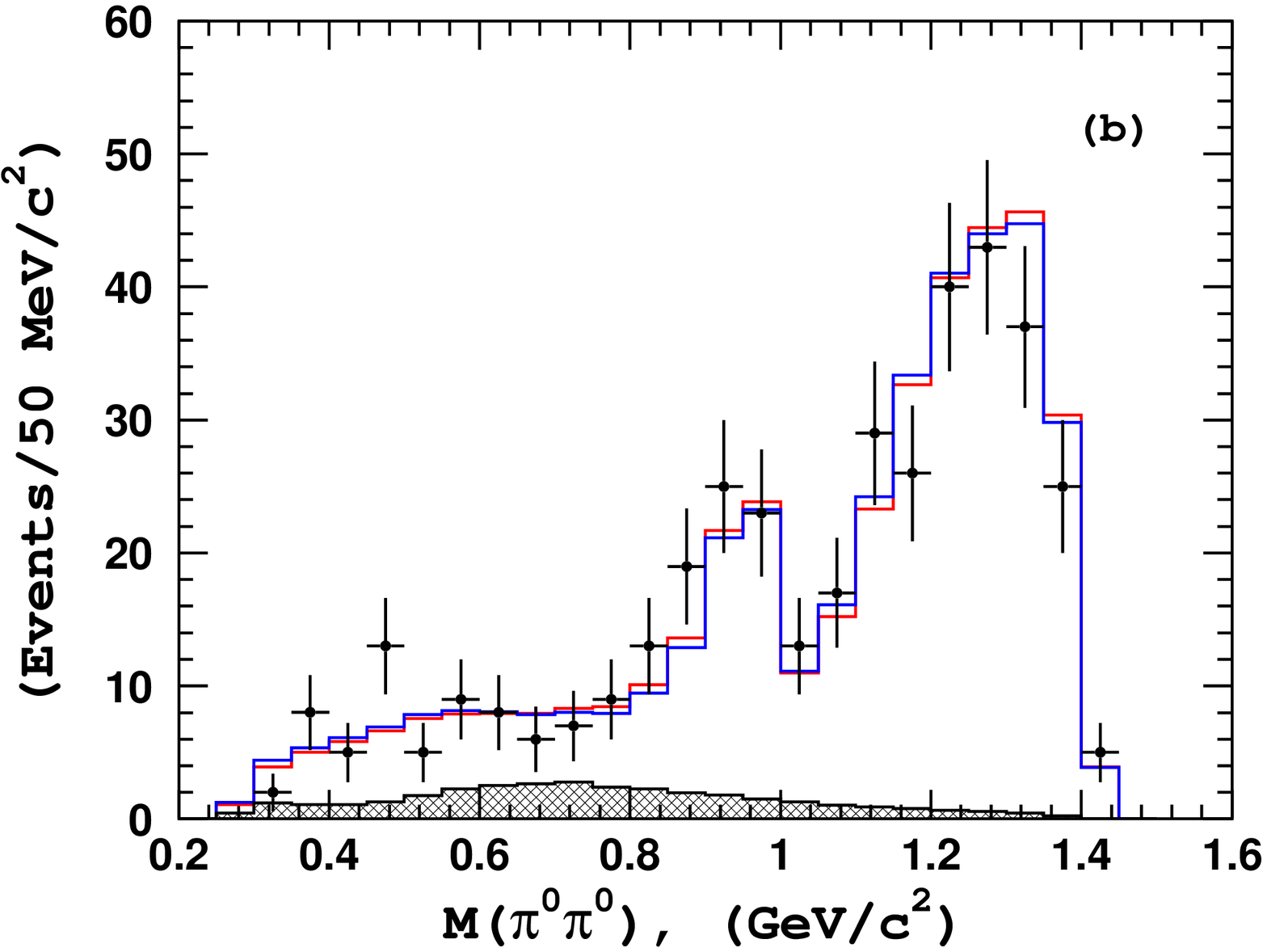}
  \includegraphics[width=0.32\textwidth] {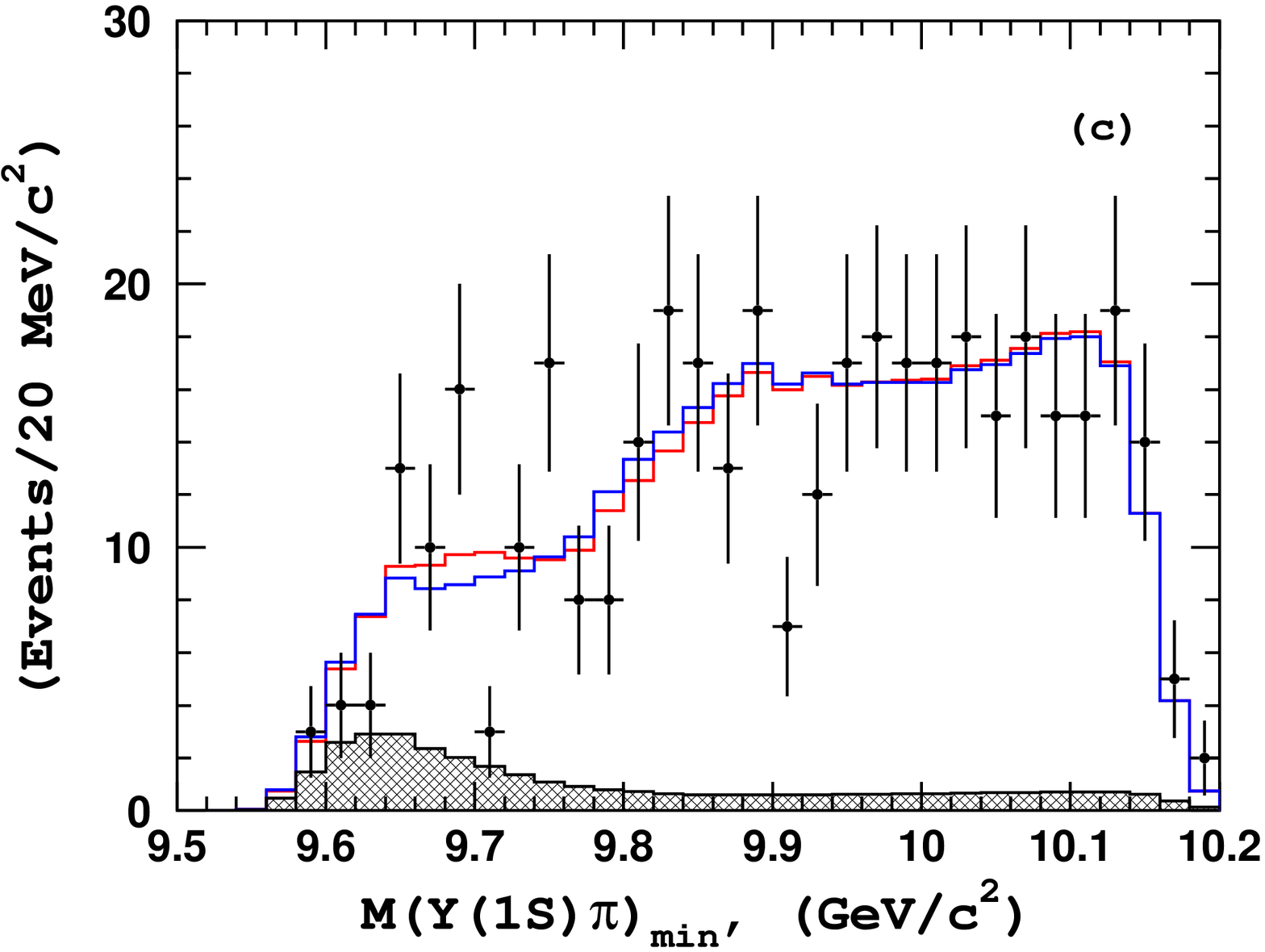}
  \caption{Comparison of the fit results (open histograms) with experimental 
    data (points with error bars) for $\up(1S)\pnpn$ events in the signal 
    region. Red and blue open histograms show the fit with and without 
    $\zbn$'s, respectively. Hatched histograms show the background components.}
  \label{fig:y1s_fitres}
\end{figure*}

\begin{figure*}
  \includegraphics[width=0.32\textwidth] {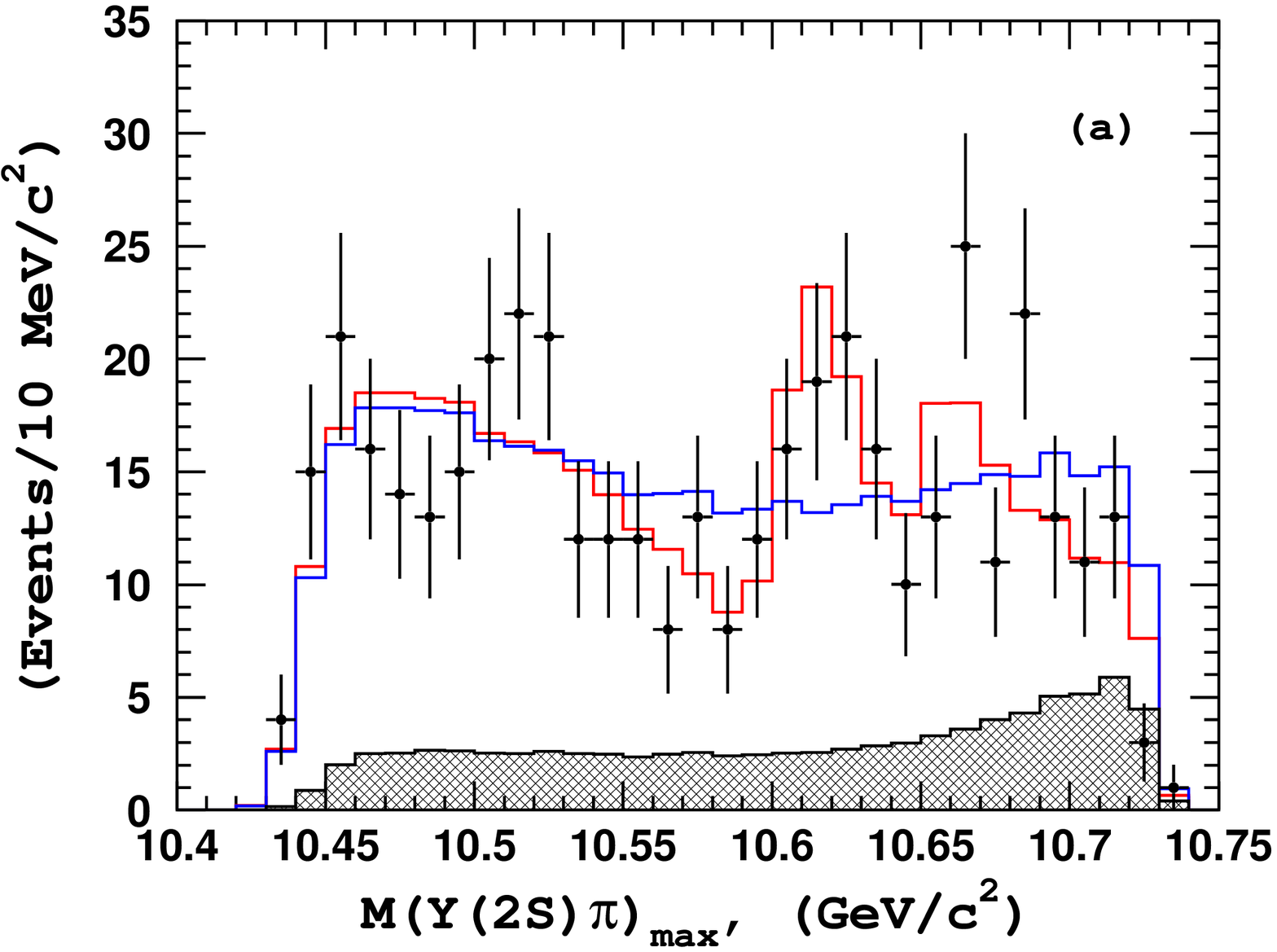}
  \includegraphics[width=0.32\textwidth] {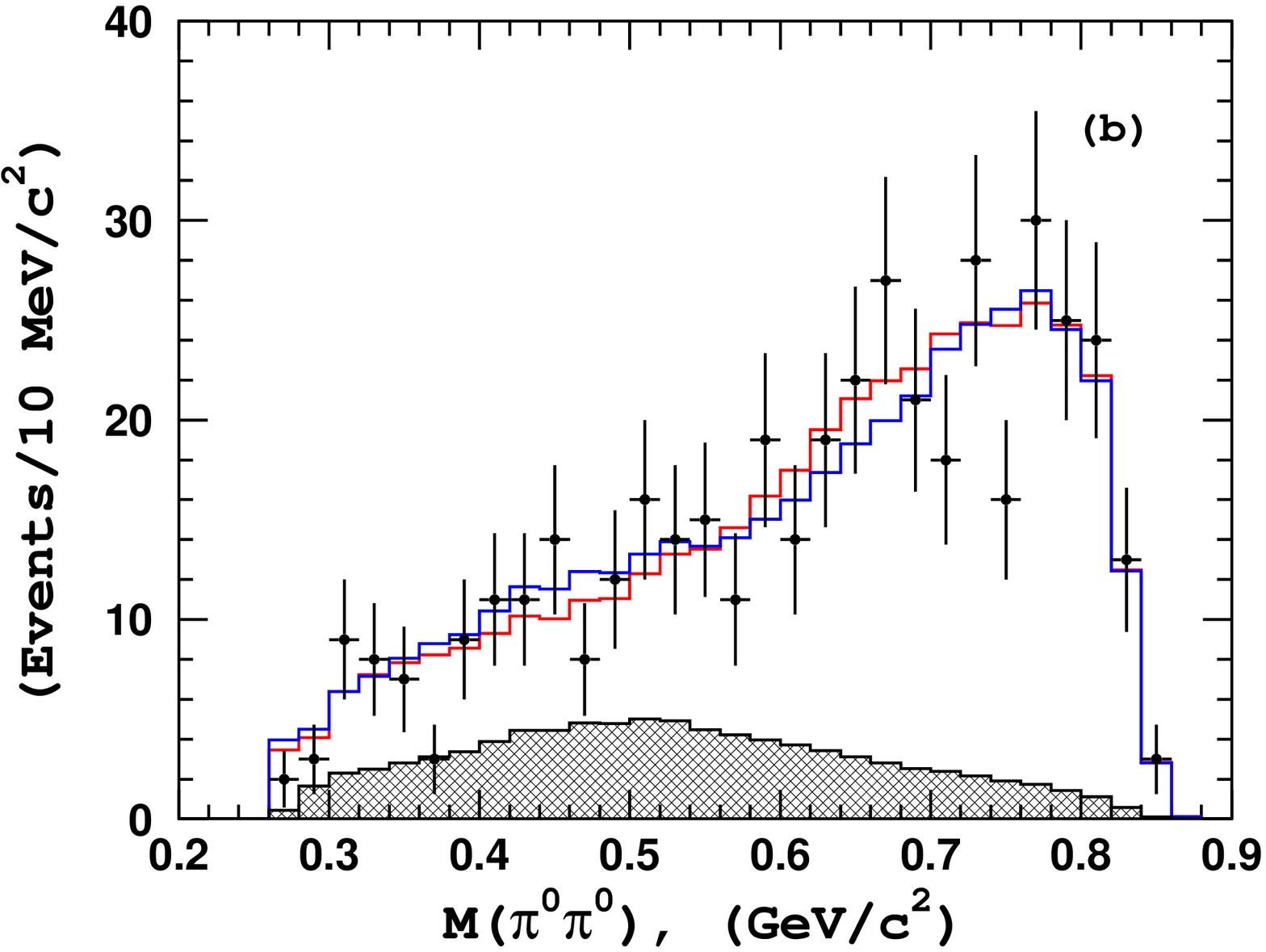}
  \includegraphics[width=0.32\textwidth] {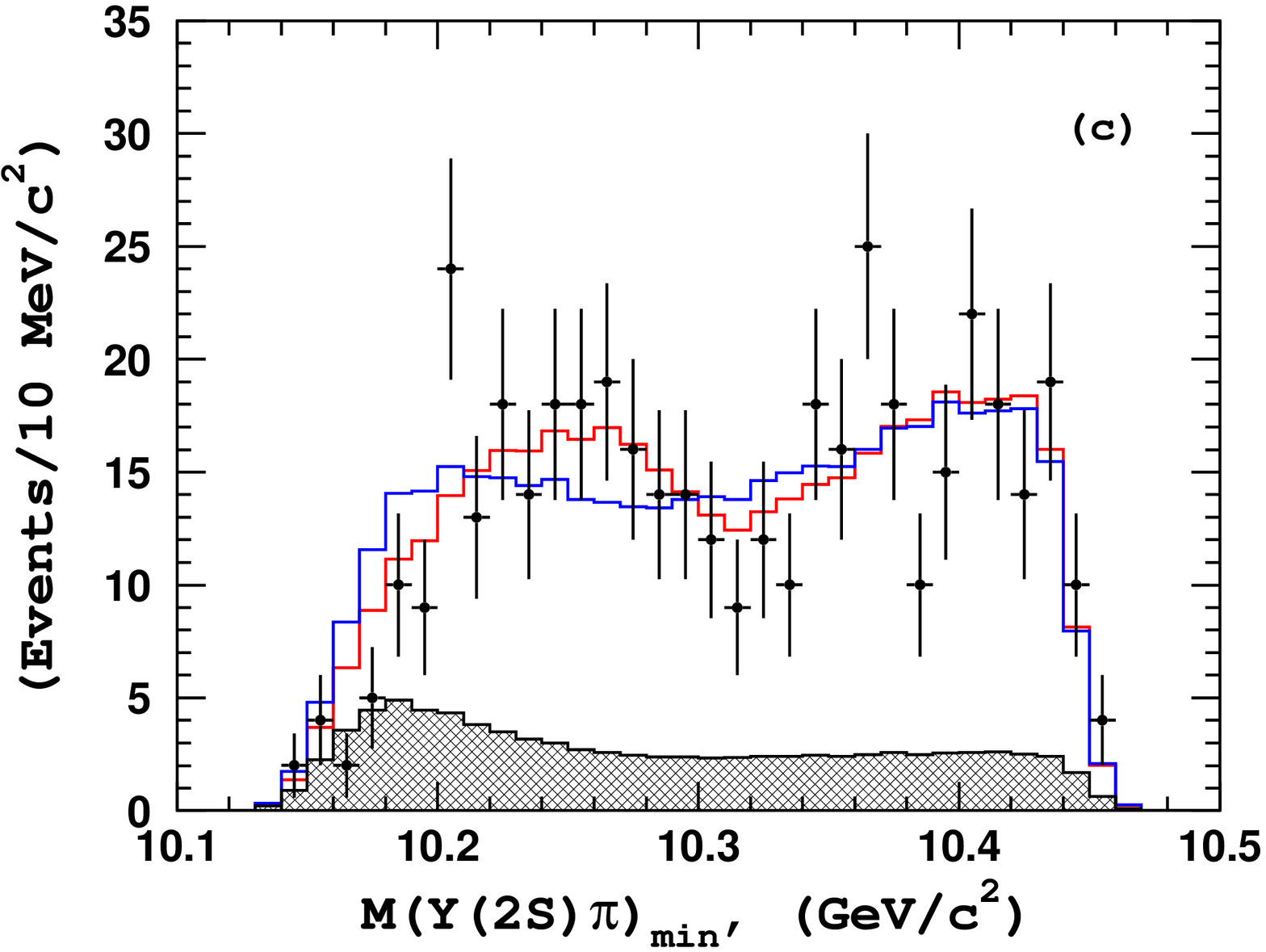}
  \caption{Comparison of the fit results (open histograms) with experimental 
    data (points with error bars) for $\up(2S)\pnpn$ events in the signal 
    region. Red and blue open histograms show the fit with and without 
    $\zbn$'s, respectively. Hatched histograms show the background components.}
  \label{fig:y2s_fitres}
\end{figure*}

Results from one-dimensional projections of the fits are shown in 
Figs.~\ref{fig:y1s_fitres} and \ref{fig:y2s_fitres}.
These projections are very similar to the corresponding distributions 
in $\upns\pipi$~\cite{zb_paper}.
A $\zbn$ signal is most clearly seen in $M(\up\pi^0)_{\rm max}$.
Table~\ref{tab:main_fit} shows the values and errors of amplitudes and 
phases obtained from the fit to the $\upns\pnpn$ Dalitz plot.
\begin{table*}
\caption{Results of the Dalitz plot fit of $\upns\pnpn$ events.
}

\medskip
\label{tab:main_fit}

\begin{tabular*}{\textwidth}{@{\extracolsep{\fill}}lccccc}\hline\hline
     & $\up(1S)\pnpn$ & $\up(1S)\pnpn$ & $\up(2S)\pnpn$ & $\up(2S)\pnpn$& $\up(2S)\pnpn$\\
Model & with $\zbn$'s & w/o $\zbn$'s   & with $\zbn$'s  &with $Z_1^0$ only & w/o $\zbn$'s\\ 
\hline
$A(Z_1^0)$ &$0.50_{-0.30}^{+0.34}$& $0.0$ \fix &$0.58_{-0.14}^{+0.21}$ &$0.47_{-0.11}^{+0.15}$ &$0.0$\fix\\
$\phi(Z_1^0)$& $-36\pm 50$   & ---        & $-113\pm 14$    & $-117\pm 17$   & ---\\
$A(Z_2^0)$& $0.60_{-0.47}^{+0.51}$& $0.0$ \fix &$0.37_{-0.16}^{+0.20}$& $0.0$ \fix  & $0.0$ \fix\\
$\phi(Z_2^0)$& $-59\pm 60$  & ---         & $-125\pm 27$    & ---            & ---\\
\hline
$A(f_2)$   & $15.7\pm 2.0$& $14.6\pm 1.6$& $18.2\pm 7.3$& $23.9\pm 7.3$ & $28.2\pm 7.0$\\
$\phi(f_2)$& $60\pm 11$   & $51\pm 9$    & $36\pm 21$  & $28\pm 13$    & $28\pm 10$\\
$A(f_0)$   &$1.07\pm 0.15$&$0.97\pm 0.12$& $11.5\pm 1.9$& $10.5\pm 1.9$& $8.2\pm 2.1$\\
$\phi(f_0)$& $168\pm 11$  & $163\pm 10$  & $211\pm 6$  & $213\pm 7$    & $210\pm  8$\\
$\anr^2$  & $15.2\pm 1.2$& $13.9\pm 0.7$&$34.7\pm 4.9$& $31.8\pm 4.3$  & $24.6\pm 4.2$\\
$\phinr^2$& $162\pm 4$  & $161\pm 4$   & $80\pm 12$  & $85\pm 13$     & $93\pm 15$\\
\hline
$-\tl$ & $-316.7$ & $-312.4$ & $-193.1$ & $-186.6$        & $-154.5$\\
\hline\hline
\end{tabular*}
\end{table*}
The statistical significance of the $\zbnf$ signal in the 
$\up(2S)\pnpn$ sample is $5.3\sigma$.
This value is obtained from the p-value, $\dtl$, with two degrees of freedom,
i.e. $\dtl=\tl({\rm w/o}\, \zbn)-\tl(10610)$.
We also perform a fit with the $\zbnf$ mass as a free parameter.
The fit with both $\zbn$'s gives $M(\zbnf)=10609^{+8}_{-6}\,\mevcc$.
A similar value, $10603\pm 6\,\mevcc$, is obtained in a fit in which only 
the $\zbnf$ is included.
The signal for the  $\zbnf$ is not significant in the fit to the 
$\up(1S)\pnpn$ events
due to the smaller relative branching fraction. The signal for the  
$\zbn(10650)$ is not significant in either $\up(1,2S)\pnpn$ dataset. 
Our data do not contradict the existence of $\zbn(10650)$, but the available statistics 
is not enough for the observation of this state.

We search for multiple solutions by doing one thousand fits with 
randomly assigned amplitudes and phases taken from a model without 
$\zbn$ contributions.
We find an additional solution in the $\up(2S)\pnpn$ final state and no other 
solutions in $\up(1S)\pnpn$. 
Table~\ref{tab:mulsol} shows the values and errors of amplitudes and phases 
obtained for both solutions. The second solution is referred to ``Solution B''.
The $\zbnf$ significance in Solution B is $5.3\sigma$, almost 
the same as in Solution A (the baseline fit).

\begin{table*}
\caption{Two solutions found in the Dalitz plot fit of $\up(2S)\pnpn$ 
events. }

\medskip
\label{tab:mulsol}

\begin{tabular*}{\textwidth}{@{\extracolsep{\fill}}lcccccc}\hline\hline
 &  w/o $\zbn$ & w/o $\zbn$ & with $Z_1^0$ & with $Z_1^0$ & with $\zbn$'s & with $\zbn$'s\\
Solutions & A & B & A & B &  A & B\\ 
\hline
$A(Z_1^0)$   & $0.0$ \fix   &  $0.0$ \fix &$0.46^{+0.15}_{-0.11}$&$1.35^{+0.64}_{-0.33}$&$0.58_{-0.14}^{+0.21}$ &$1.42\pm 0.48$\\
$\phi(Z_1^0)$& ---          &  ---          & $-117\pm 14$  & $88\pm 18$     & $-113\pm 14$     & $91\pm 21$ \\
$A(Z_2^0)$   & $0.0$ \fix   &  $0.0$ \fix   & $0.0$ \fix    &  $0.0$ \fix      &$0.37_{-0.16}^{+0.20}$ & $0.66\pm 0.40$\\
$\phi(Z_2^0)$& ---          &  ---          & ---           &  ---             & $-125\pm 27$     & $124\pm 37$ \\
\hline
$A(f_2)$    & $28.2\pm 7.0$& $41.8\pm 9.0$ & $23.9\pm 7.3$ & $48.7\pm 15.4$   & $18.2\pm 7.3$     & $43.3\pm 15.6$\\
$\phi(f_2)$ & $28\pm 10$   & $-1\pm 14$    & $18\pm 13$    & $10\pm 16$       & $36\pm 21$        & $132\pm 19$\\
$A(f_0)$    & $8.2\pm 2.1$ & $13.3\pm 3.6$ & $10.5\pm 1.9$ & $13.4\pm 4.2$    & $11.5\pm 1.9$     & $12.6\pm 4.9$\\
$\phi(f_0)$ & $210\pm  8$  & $131\pm 11$   & $213\pm 7$    & $134\pm 15$      & $211\pm 6$        & $132\pm 19$    \\
$\anr^2$     & $24.6\pm 4.2$& $44.2\pm 10.1$& $31.8\pm 4.3$ & $50.4\pm 12.2$  & $34.7\pm 4.9$     & $50.8\pm 13.7$ \\
$\phinr^2$  & $93\pm 15$   & $-70\pm 16$   & $85\pm 13$    & $-69\pm 22$      & $80\pm 12$        & $-72\pm 25$ \\
\hline
$-\tl$      & $-154.5$     & $-155.4$      & $-186.6$       & $-186.3$        & $-193.1$          & $-191.2$  \\
\hline\hline
\end{tabular*}
\end{table*}

\section{Study of systematic uncertainties in Dalitz Analysis}
\label{sec:syst}

Experimental errors may arise from the uncertainty in parameterization of 
the background PDF.
We determine this uncertainty by varying parameters of the background PDF.
We use different sideband sub-samples to determine PDF parameters:
the low-mass sideband only, or the high-mass sideband, or 
$\upns\to\ee$ events only, or $\upns\to\mumu$ events only.
The statistical significance of the $\zbnf$ in all fits is greater than 
$4.9\sigma$.
Another source of systematic uncertainty is the determination of signal 
efficiency. To estimate this effect we perform two additional fits with 
a modified efficiency function:
$\sqrt{\epsilon(s_{\rm max},s_3)}$ and $\epsilon^{3/2}(s_{\rm max},s_3)$.
The result, $\dtl$ for models with $\zbnf$ and without $\zbn$'s 
changes from $32.1$ to $32.9$ and $31.2$, respectively.
the difference in $\zbnf$ significance is less than $0.1$.
We also perform a fit with a modified detector resolution function:
the resolutions are varied from $4$ to $8\,\mevcc$ instead of the nominal 
$6\,\mevcc$ to take into account the effect of different momentum 
resolutions in MC  and data.
The resulting changes in $\dtl$ are less than $0.5$ 
(difference in $\zbnf$ significance is less than $0.05$).

The model uncertainty originates mainly due to the parameterization 
of the non-resonant 
amplitude. To estimate it we vary the model used to fit the data. 
Three additional models, based on solution A, are used:
with an additional $\sigma(600)$ resonance, parameterized by a Breit-Wigner 
function with $M=600\,\mevcc$ and $\Gamma=400$~MeV$/c$,
a model with $\anr=a e^{i\phi_a} + b e^{i\phi_b} \sqrt{s(\pnpn)}$,
and a model without any $f_0(980)$ contribution 
(to fit $\up(2S)\pnpn$ events only). 
The smallest $\zbnf$ significance is obtained in the last model: $4.9\sigma$. 
We use this value as the final $\zbnf$ significance.
Fits with the $\zbnf$ mass as a free parameters give values from
$10603$ to $10615\,\mevcc$. We use $\pm 6\,\mevcc$ as a model uncertainty for 
the $\zbnf$ mass.

\section{Conclusion}
We report the observation of $\upf\to\up(1,2S)\pnpn$ decays.
The measured branching fractions,
$\br(\upf\to\up(1S)\pnpn)=(2.25\pm 0.11\pm 0.20)\times 10^{-3}$,
$\br(\upf\to\up(2S)\pnpn)=(3.66\pm 0.22\pm 0.48)\times 10^{-3}$,
are found to be consistent 
with the the expectation from isospin, scaling from 
$\br(\upf\to\up(nS)\pipi)$~\cite{belle_ypipi}.

Evidence of a neutral resonance decaying to $\up(2S)\pi^0$, $\zbnf$, 
has been obtained in a Dalitz plot analysis of $\upf\to\up(2S)\pnpn$ decay.
The statistical significance of the $\zbnf$ signal is $5.3\sigma$ 
($4.9\sigma$ including model and systematic uncertainties).
Its measured mass, $M(\zbnf)=10609_{-6}^{+8}\pm 6\,\mevcc$, is
consistent with the mass of the corresponding charged state, the $\zb^\pm(10610)$.
The $\zbn(10650)$ signal is not significant in either $\up(1S)\pnpn$ or
$\up(2S)\pnpn$. 
Our data do not contradict the existence of $\zbn(10650)$, but the 
available statistics is not enough for the observation of this state.

\section{ Acknowledgments}
We thank the KEKB group for the excellent operation of the
accelerator; the KEK cryogenics group for the efficient
operation of the solenoid; and the KEK computer group,
the National Institute of Informatics, and the 
PNNL/EMSL computing group for valuable computing
and SINET4 network support.  We acknowledge support from
the Ministry of Education, Culture, Sports, Science, and
Technology (MEXT) of Japan, the Japan Society for the 
Promotion of Science (JSPS), and the Tau-Lepton Physics 
Research Center of Nagoya University; 
the Australian Research Council and the Australian 
Department of Industry, Innovation, Science and Research;
the National Natural Science Foundation of China under
contract No.~10575109, 10775142, 10875115 and 10825524; 
the Ministry of Education, Youth and Sports of the Czech 
Republic under contract No.~LA10033 and MSM0021620859;
the Department of Science and Technology of India; 
the Istituto Nazionale di Fisica Nucleare of Italy; 
the BK21 and WCU program of the Ministry Education Science and
Technology, National Research Foundation of Korea,
and GSDC of the Korea Institute of Science and Technology Information;
the Polish Ministry of Science and Higher Education;
the Ministry of Education and Science of the Russian
Federation and the Russian Federal Agency for Atomic Energy;
the Slovenian Research Agency;  the Swiss
National Science Foundation; the National Science Council
and the Ministry of Education of Taiwan; and the U.S.\
Department of Energy and the National Science Foundation.
This work is supported by a Grant-in-Aid from MEXT for 
Science Research in a Priority Area (``New Development of 
Flavor Physics''), and from JSPS for Creative Scientific 
Research (``Evolution of Tau-lepton Physics'').


\begin{thebibliography}{99}

\bibitem{zb_paper}
A. Bondar, A. Garmash, R. Mizuk, D. Santel, K. Kinoshita, et al. (Belle Collaboration)
Phys. Rev. Lett. 108, 122001 (2012).

\bibitem{zb_helicity}
I. Adachi et al. (Belle Collaboration), arXiv:1105.4583.

\bibitem{zb_molecular}
A.E. Bondar, A. Garmash, A.I. Milstein, R. Mizuk, M.B. Voloshin,
Phys. Rev. D 84, 054010 (2011).

\bibitem{belle}
A. Abashian et al. (Belle Collaboration), 
Nucl. Instrum. Methods Phys. Res., Sect. A 479, 117 (2002).

\bibitem{kekb}
S. Kurokawa and E. Kikutani, 
Nucl. Instrum. Methods Phys. Res. Sect. A 499, 1 (2003), 
and other papers included in this Volume.

\bibitem{kpipi}
A. Garmash et al. (Belle Collaboration), Phys.Rev.Lett. 96, 251803 (2006).

\bibitem{PDG}
J. Beringer et al. (Particle Data Group), Phys. Rev. D86, 010001 (2012).

\bibitem{voloshin}
M.B. Voloshin, Prog. Part. Nucl. Phys. 61, 455 (2008).
M.B. Voloshin, Phys. Rev. D 74, 054022 (2006) and references therein.

\bibitem{belle_ypipi}
K.-F. Chen, W.-S. Hou, M. Shapkin, A. Sokolov et al. (The Belle collaboration)
Phys. Rev. Lett. 100 112001 (2008).

\end{thebibliography}
\end{document}